\documentstyle[12pt]{article}
\def\be{\begin{equation}}
\def\ee{\end{equation}}
\def\ba{\begin{eqnarray}}
\def\ea{\end{eqnarray}}

\def\fun#1#2{\lower3.6pt\vbox{\baselineskip0pt\lineskip.9pt

\ialign{$\mathsurround=0pt#1\hfill##\hfil$\crcr#2\crcr\sim\crcr}}}
\begin{document}

\begin{titlepage}
\begin{flushright}
UMD-PP-95-142 \\
hep-ph/9507239 \\
\end{flushright}
\null\vspace{-62pt}
\vspace{3.0cm}
\centerline{{\large \bf  Many Higgs Doublet Supersymmetric Model, }}
\centerline{{\large \bf  Flavor Changing Interactions }}
\centerline{{\large \bf  and Spontanoeus CP Violation$^*$ }}
\vspace{0.5in}
\centerline{Andrija Ra\v{s}in}
\centerline{{\it Department of Physics,
University of Maryland, College Park, MD 20742, USA}}
\vspace{.05in}
\baselineskip=22pt

\vspace{3cm}

\centerline{\bf Abstract}
\begin{quotation}
I study many higgs doublet supersymmetric model with spontaneous
CP violation. The damaging flavor changing interactions and large
CP violation are brought under control simultaneously by an
approximate PQ symmetry. I relate the smallness of the CP violating
parameter in the kaon system to the small
${m_b \over m_t}$ ratio.

\vspace{1cm}

\noindent $^*$ Talk given at the XXXth Rencontres De moriond, ``Electroweak
Interactions and Unified Theories", Les Arcs, France, March 11-18, 1995.

\end{quotation}
\end{titlepage}

\baselineskip=21pt

Today the standard model is satisfactorily describing
vast amount of data, but many free input parameters that the
theory has makes it hardly anyone believe to be
the ultimate theory. Especially interesting is the sector
of CP violation and fermion masses, which carries the most
of the unknown parameters.
CP violation can be acommodated in the standard model through
a complex phase in the Kobayashi-Maskawa (KM) matrix,
but a theoretically more promising idea is
spontaneous CP symmetry violation (SCPV)$^{1)}$.
SCPV has been used also in the
supersymmetric extensions of the
standard model (SSM). It was studied
in SSM with the addition of an singlet
(see for example Ref. 2).

In this talk I consider the SCPV
in the extension of the SSM with more higgs doublets.
The four higgs doublet SSM (at least two doublets must be
added to the MSSM in order to make the theory anomaly free)
occurs naturally in left-right supersymmetric models where at
least two bidoublets are required in order to get realistic
fermion masses and mixings. Also it was shown$^{3)}$ to be
the simplest extension that has naturally large tan$\beta$
(the ratio of the vevs of the doublets coupled to up and down
quark sectors)
without fine tuning the theory.

Unless some terms in the lagrangian are forbidden by some additional
symmetries, the many higgs doublet SSM will
in principle have flavor changing interactions (FC).
For vevs and masses of the new higgs scalars of the order of
weak scale or so these are below or of the order of experimental
limits if the Yukawa couplings are real and are small by some
approximate flavor symmetry mechanism$^{4)}$.
However, allowing for complex Yukawa couplings in the
Lagrangian, the amount of CP violation is by many orders
of magnitude larger than it is observed$^{5)}$.
Also, allowing for large ratio of vevs the FC interactions may be
too big. The purpose of this talk is to show that
by using an approximate Peccei-Quinn type symmetry (PQ) we can bring
both problems (too large FC and too large CP) under control.

The most general superpotential with four higgs doublets
$H_i,i=1,2,3,4$ (for previous work on four higgs doublets in
supersymmetry see Refs. 6 and 7) is given by
\ba
W & = & Q ( h_1 H_1 + h_3 H_3) D^c + Q ( h_2 H_2 + h_4 H_4) U^c +
        L(h_1^e H_1 + h_3^e H_3) E^c \nonumber\\
& + & \mu_{12} H_1 H_2 + \mu_{32} H_3 H_2
+ \mu_{14} H_1 H_4 + \mu_{34} H_3 H_4,
\label{eq:superpot}
\ea
where $H_1,H_3$ have hypercharge $-1$, and
$H_2,H_4$ have hypercharge $+1$. $h_i$ are the Yukawa matrices.

The most general scalar potential is given by
\ba
V & = & m_1^2 H_1^\dagger H_1+ m_2^2 H_2^\dagger H_2
 + m_3^2 H_3^\dagger H_3
+ m_4^2 H_4^\dagger H_4 - \nonumber\\
 & - & (m^2_{12} H_1 H_2 + h.c.) - (m^2_{32} H_3 H_2 + h.c.) - \nonumber\\
 & - & (m^2_{14} H_1 H_4 + h.c.) - (m^2_{34} H_3 H_4 + h.c.) - \nonumber\\
 & - & (m^2_{13} H_1^\dagger H_3 + h.c.)
- (m^2_{24} H_2^\dagger H_4 + h.c.) + V^{4HD}_D ,
\label{eq:pot}
\ea
where $V^{4HD}_D$ is the D-term part of the potential.
Unless the parameters are suppressed by some symmetry,
we will assume that all the dimensionfull paramaters
are of the order of the weak scale, while all dimensionless parameters
are of order one. We do not assume any higher energy scales
or accidental cancellations.
We will also assume that the theory is CP invariant, i.e.
all the couplings are real.
The neutral components
of Higgs fields will acquire complex vacuum expectation
values
$<H_1> = {v_1 \over \sqrt{2} } $,
$<H_3> = {v_3 e^{i\delta_3} \over \sqrt{2} }$
and
$<H_2> = {v_2 e^{i\delta_2} \over \sqrt{2} }$,
$<H_4> = {v_4 e^{i\delta_4} \over \sqrt{2} }$,
where we rotated away the trivial phase of $<H_1>$.
The vacuum expectation value of the scalar potential is
\ba
<V> & = &{1 \over 2} m_1^2 v_1^2 +
{1 \over 2} m_2^2 v_2^2 +
{1 \over 2} m_3^2 v_3^2 +
{1 \over 2} m_4^2 v_4^2 -
 m_{32}^2 v_3v_2 \cos(\delta_3+\delta_2)
- m_{14}^2 v_1 v_4 \cos \delta_4 \nonumber \\
& - & m_{12}^2 v_1 v_2 \cos \delta_2
- m_{13}^2 v_1 v_3 \cos \delta_3
- m_{34}^2 v_3 v_4 \cos(\delta_3+\delta_4)
- m_{24}^2 v_2 v_4 \cos(\delta_2-\delta_4) + \nonumber \\
& + & {1 \over 32} (g^2 + g'^2) [ v_1^2 + v_3^2 - v_2^2 - v_4^2 ]^2 .
\label{eq:vevpot}
\ea
The necessary phase conditions at the minimum are
\ba
\frac {\partial V} {\partial \delta_2} & = &
    m_{12}^2 v_1 v_2 \sin \delta_2
    + m_{32}^2 v_3 v_2 \sin (\delta_3 + \delta_2)
    + m_{24}^2 v_2 v_4 \sin (\delta_2 - \delta_4) = 0 , \nonumber\\
\frac {\partial V} {\partial \delta_3} & = &
     m_{32}^2 v_3 v_2 \sin (\delta_3 + \delta_2)
    + m_{13}^2 v_1 v_3 \sin \delta_3
    + m_{34}^2 v_3 v_4 \sin (\delta_3 + \delta_4) = 0 , \nonumber\\
\frac {\partial V} {\partial \delta_4} & = &
    - m_{24}^2 v_2 v_4 \sin (\delta_2 - \delta_4)
    + m_{34}^2 v_3 v_4 \sin (\delta_3 + \delta_4)
    + m_{14}^2 v_1 v_4 \sin \delta_4 = 0 .
\label{eq:minphases}
\ea
It is
plausible that a general solution with the values
for phases different from zero or $\pi$
can be found.
For coefficients $m$ of the order of weak scale or so,
we expect the vevs to be of the same order, as well as the phases
of the vevs to be of order one.
However, in the absence of any additional symmetries, many
higgs doublet models will have flavor changing
interactions. This is because diagonalization of
the Yukawa matrix coupled to one Higgs will not
in general diagonalize the Yukawa matrix of the other
Higgs (otherwise there is at least a discrete symmetry
which relates the two matrices).

\vspace{0.22in}

{\bf Yukawa couplings, FC and the
Peccei-Quinn like symmetry}

The limits on FC can be avoided if we introduce some symmetry
(or approximate symmetry) that will
decouple (or almost decouple) the second pair of higgses from fermions.
In this way the FC will be proportional to the amount by which the symmetry
is broken.

A simple scenario for the couplings in a four higgs
doublet SSM is to impose an approximate symmetry
which is very similar to Peccei-Quinn symmetry (I call
it the PQ symmetry). The terms that violate the symmetry get suppressed
by powers of a small factor $\epsilon_{PQ}$.
I assume following assignment of the charges:
$ Q(H_3)  =  +1$,$Q(H_4)= -1$ and $Q(D^c) = +1$.
All other fields have zero PQ charge.
This is similar to
the model of Nelson and Randall$^{3)}$. The only difference
is that in their model $D^C$ has charge -1.

{}From the assignments of the PQ charges we observe
that in the superpotential (\ref{eq:superpot}) $h_3$
is suppressed by $\epsilon_{PQ}^2$,
$h_1$ and $h_4$ are suppressed by $\epsilon_{PQ}$,
while $h_2$ is unsuppressed and similarly for $\mu_{ij}$.
Same as in Ref. 3, we will
take $\epsilon_{PQ} = \frac {1} {\tan\beta} = { m_b \over m_t}$.
This gives the explanation of the large ratio of $t$ to $b$ mass
entirely in terms of the approximate PQ symmetry, while the
hierarchy between the generations of the same charge
is left to the flavor symmetry
breaking part.
The assignments of charges also tells us that
in the scalar potential $m_{14}^2,m_{32}^2,m_{13}^2,m_{24}^4$ are
suppressed by $\epsilon_{PQ}$. $m_{12}^2$ and $m_{34}^2$ remain
unsuppressed (order weak scale).

Depending on the
choice of the parameters, the minimum of the potential
will be when one and only one pair of vevs $(v_1,v_2)$ or $(v_3,v_4)$
is suppressed by $\epsilon_{PQ}$ (compared to the weak scale).
Otherwise we get a higher minimum,
or an unbounded potential. This is obvious from looking
at the vev minimum conditions for the potential.
We choose the pair $(v_3,v_4)$ to be suppressed,
while $v_1$ and $v_2$ remain unsuppressed. In this way light
goldstone bosons
are avoided since the size of breaking is of the same order as the
explicit symmetry breaking terms in the Lagrangian$^{3)}$.
Notice that the whole effect of the change of the PQ assignment
of $D^c$ is that the Yukawa of $H_3$ is suppressed by
$\epsilon_{PQ}^2$. Since the Yukawa of $H_1$, which is primarily of the order
of the down quark mass
matrix, is down by $\epsilon_{PQ}$, this means that FC couplings
will have an additional factor of $\epsilon_{PQ}$.
Notice that the assignment of charge $+1$ to $D^c$
was crucial. If it were $-1$, the yukawa couplings $h_1$ would not
have had any PQ suppression, and the theory would have damaging
FC.
The authors of Ref. 3 introduce the additional assumption
of a spurion field in order to avoid couplings of $H_1$,
thus explaining the large ratio ${m_t \over m_b}$ with large
ratio of vevs $\tan\beta$.
With our assignment of the charges the large ratio
${m_t \over m_b}$ is explained by different suppressions of the Yukawa
matrices. We do {\it not} have large $\tan\beta$, but the origin
of large ${m_t \over m_b}$ is the same, namely approximate PQ symmetry.

Next, we can see from (\ref{eq:minphases})
that only $\sin\delta_2$ must be
suppressed by $\epsilon_{PQ}^2$, while other phases
can be of order one. This is our general result:
by allowing for an approximate symmetry,
some phases which were CP trivial (i.e. $0$ or $\pi$)
in the limit of exact symmetry,
may become of order one. Crucial is that these phases
are {\it not} proportional to $\epsilon_{PQ}$.

Notice that the elements of the KM matrix will be real
up to the leading order in $\epsilon_{PQ}$,
since the up and down mass matrices
couple dominantly to one higgs only,
namely to $H_1$ and $H_2$. The phases of the vevs of $H_1$
and $H_2$ do not enter the diagonalization matrices
and thus do not enter the KM matrix.
We can compute the contribution of the exchange
of flavor changing scalars to
$\Delta M_K$.  It will be down by $\epsilon_{PQ}^2$ (two couplings)
compared to having just couplings comparable to those
responsible for quark masses.
For scalars of order weak scale or so, this means that the
standard model box diagram will be the dominant contribution
to $\Delta M_K$. However this contribution is real since the
KM matrix is real.
Thus, the dominant contribution to the CP violating parameter
$\epsilon_{CP}$ will come from the flavor changing scalar exchange.
Although the scalars contribute a phase of order one its
amplitude is suppressed by $\epsilon_{PQ}^2$, and this is what
makes $\epsilon_{CP}$ small.
The contribution of the flavor mediating scalars
is of the same order as the standard model contribution when
the mass of the scalars is about 1 TeV or so.
Thus,
\be
\epsilon_{CP} \approx \frac { Im \Delta M_K} {\Delta M_K }
\approx \epsilon_{PQ}^2 (\frac {1 TeV} {M})^2 \sin\phi
\approx 3 \times 10^{-3} ( {{ m_b \over m_t }\over { 1 \over 60 }})^2
(\frac {300 GeV} {M})^2 \sin\phi,
\ee
where M is the typical mass of the flavor mediating Higgs scalar,
and $\sin\phi$ is a CP violating phase of order one.
This actually gives the right value
for $M=300GeV$, which is the weak scale (we took $m_b = 3 GeV$ at the
weak scale)! We do not need scalars of order
TeV, which Hall and Weinberg considered to be somewhat heavy
anyway.

The suppression of CP violating effects because of
the approximate PQ symmetry despite the existence of phases
of order one is not a property only of the kaon system.
Any flavor changing exchange is necessarily suppressed by
powers of $\epsilon_{PQ}$. The contributions to different
diagrams will come either from the large vev of $H_2$,
which has a suppressed phase, or from other doublets
which either have suppressed vevs or small Yukawa couplings.
For this reason we expect direct CP violating effects
will also be very small, as well as the CP violating
phases in the B system. However we expect the neutron electric
dipole moment (NEDM) to be of the order of experimental limit.
This is because naive estimates which do not include
any suppression of the flavor changing couplings usually
give NEDM several orders of magnitude higher than the
experimental limit.
This general situation is similar to the
assumption of suppressed CP violating phases of
Hall and Weinberg$^{5)}$. Here we offer an explanation
for the smallness of CP violation in terms of the small
ratio of bottom and top quark masses and we link it to
the PQ symmetry.

We also note that as far as chargino masses
go we have no light charginos$^{3)}$, because the
PQ numbers for higgses are the same as in Ref. 3.

Finally, no attempt was made to include this scheme into
a grand unified theory. However, many attempts
have higgs multiplets which have more than two doublets
(for example two 10's in SO(10) or 2 bidoublets in
a LR model). Whether these doublets can stay light
will depend on the details of the theory, such as
intermediate scales.

I thank the organizers of the Moriond conference
for invitation and support.
I would also like to thank Jongbae Kim,
Rabi Mohapatra, Alex Pomarol and Lisa Randall for useful
discussions, and Jo\~{a}o Silva who participated in the early stages
of the project. This work was supported by the NSF grant PHY9421385.

{\bf Note: } After the talk, we made more progress on understanding what
the conditions for SCPV in many higgs
SUSY model are. It appears that the tree level
potential is not sufficient by itself, and an additional contribution
(radiative corrections, soft CP violation or an additional singlet)
is needed$^{8)}$.

{\bf References}

\noindent 1. T. D. Lee, Phys. Rev. {\bf D8}, 1226 (1973).

\noindent 2. A. Pomarol, Phys. Rev. {\bf D47}, 273 (1993);
K. S. Babu and S. M. Barr, Phys. Rev. {\bf D49}, 2156 (1994).

\noindent 3. A. E. Nelson and L. Randall, Phys. Lett. {\bf B316},
516 (1993).

\noindent 4. C. D. Froggatt and H. B. Nielsen, Nucl. Phys.
{\bf B147}, 277 (1979); T. P. Cheng and M. Sher, Phys. Rev. {\bf D35},
3484 (1987);
A. Antaramian, L. J. Hall and A. Ra\v{s}in, Phys. Rev. Lett. {\bf 69},
1871 (1992).

\noindent 5. L. Hall and S. Weinberg,
Phys. Rev. {\bf D48}, 979 (1993).

\noindent 6. H. Haber and Y. Nir, Nucl. Phys. {\bf B335}, 363, (1990).

\noindent 7. R. A. Flores and M. Sher,
Ann. Phys. {\bf 148}, 95 (1983); K. Griest and M. Sher, Phys. Rev.
{\bf D42}, 3834, (1990).

\noindent 8. M. Masip and A. Ra\v{s}in, in preparation.

\end{document}